\documentclass[pdflatex,sn-mathphys-num]{sn-jnl}

\usepackage{graphicx}%
\usepackage{multirow}%
\usepackage{amsmath,amssymb,amsfonts}%
\usepackage{amsthm}%
\usepackage{mathrsfs}%
\usepackage[title]{appendix}%
\usepackage{xcolor}%
\usepackage{textcomp}%
\usepackage{manyfoot}%
\usepackage{booktabs}%
\usepackage{algorithm}%
\usepackage{algorithmicx}%
\usepackage{algpseudocode}%
\usepackage{listings}%

\theoremstyle{thmstyleone}%
\theoremstyle{thmstyletwo}%

\theoremstyle{thmstylethree}%

\begin{document}

\title[A new Fractal Mean-Field analysis in phase transition]{A new Fractal Mean-Field analysis in phase transition}


\author*[1,2]{\fnm{Ismael} \sur{S. S. Carrasco}}\email{ismael.carrasco@unb.br}

\author[1]{\fnm{Henrique} \sur{A. de Lima}}\email{henrique\_adl@hotmail.com}
\equalcont{These authors contributed equally to this work.}

\author[1]{\fnm{Fernando} \sur{A. Oliveira}}\email{faooliveira@gmail.com}
\equalcont{These authors contributed equally to this work.}

\affil*[1]{\orgdiv{International Center of Physics, Institute of Physics}, \orgname{University of Brasilia}, \orgaddress{\city{Brasilia}, \postcode{70910-900}, \state{Federal District}, \country{Brazil}}}

\affil*[2]{\orgdiv{Departamento de F\'isica}, \orgname{Universidade Federal de Vi\c cosa}, \orgaddress{\city{Vi\c cosa}, \postcode{36570-900}, \state{MG}, \country{Brazil}}}


\abstract{Understanding phase transitions requires not only identifying order parameters but also characterizing how their correlations behave across scales. By quantifying how fluctuations at distinct spatial or temporal points are related, correlation functions reveal the structural organization of complex systems. In this work, we reexamine the theoretical foundations of these correlations in systems undergoing second-order phase transitions, with emphasis on the Ising model extended to non-integer spatial dimensions. We revisit the hypotesis that, at criticality, the equilibrium dynamics become effectively confined to the fractal edge of spin clusters and redo the analysis using fractional calculus. Within this framework, the fractal dimension that governs the correlations in that subspace is directly related to Fisher’s exponent $\eta$, which quantifies the singular behavior of the correlation function near criticality. Importantly, this correlation fractal dimension is distinct from the fractal dimension associated with the order parameter. The fractional approach allows us to directly compute the correlation fractal dimension and to establish an explicit geometrical relation connecting the two fractal dimensions. Moreover, the formulation naturally extends to non-integer spatial dimensions, remaining valid below the upper critical dimension and yielding the correct value of Fisher’s exponent $\eta$ for a continuous spatial dimension $d$. Within this framework, we also provide empirical functions describing how the main critical exponents vary continuously as a function of the space dimension.
}


\maketitle

\section{Introduction}
\label{Int}
Collective reorganization of microscopic degrees of freedom underlies the dramatic macroscopic changes observed near critical points. Such critical behavior cannot be fully captured by single-site observables alone: it is the spatial and temporal patterns of correlations that encode the emergent order and the breakdown of symmetry across scales \cite{Kardar07}. In many systems of interest, from magnetic materials to growing interfaces, these rearrangements manifest themselves through scale-invariant structures that require geometric as well as statistical descriptions.

The language of fractals has proven particularly fruitful in this context. Since Mandelbrot's work \cite{Mandelbrot82}, fractal geometry has been recognized as a natural language for the self-similar patterns that appear near transitions. Indeed, the renormalization and scaling properties, central to critical phenomena, are closely linked to the underlying fractal organization~\cite{Cardy96,Suzuki8,Kroger00,Devakul19}. Thus, geometric characterizations complement the usual thermodynamic viewpoint and help clarify both the successes and the limits of mean-field approaches~\cite{Lima24,Lima25}.

The autocorrelation function is a central observable for this discussion, given by
\begin{equation}
	\label{G0}
	G(r)=\langle \psi(\vec{r}+\vec{x})\psi(\vec{x})\rangle,
\end{equation}
with $\psi(\vec{x})$ the fluctuation of the local order-parameters and $\langle\cdots\rangle$ the equilibrium average in a $d$-dimensional lattice. The Ornstein--Zernike approximation \cite{Ornstein14,Kardar07} leads, in homogeneous continuous media, to the formal relation
\begin{equation}
	\label{G}
	(-\nabla^2 +\rho^{-2})G(r)=\delta^{(d)}(r)  \ . 
\end{equation} 

Its solution, $G(r)\propto r^{2-d}\mathrm{e}^{-r/\rho}$, captures the large-scale correlation decrease, with $G(r)\propto r^{2-d}$ at the critical point where $\rho\to\infty$, the correlation length diverging as $\rho\propto|T-T_c|^{-\nu}$. However, this behavior breaks down at certain dimensions: for $d=2$ it predicts a constant correlation, while for $d<2$ it yields a non-physical increase with distance. Empirically, scattering experiments contradict this mean-field prediction, prompting Fisher's introduction of the exponent $\eta$, which follows the scaling relation
\begin{equation}
	\label{gamma}
	\gamma=(2-\eta)\nu,
\end{equation}
and corrects the short-distance scaling of $G(r)$ \cite{Fisher64}:
\begin{equation}
	\label{G2}
	G(r) \propto
	\begin{cases}
		r^{2-d}\,\mathrm{e}^{-r/\rho}, & \text{if } r>\rho,\\[4pt]
		r^{2-d-\eta}, & \text{if } r\ll\rho.
	\end{cases}
\end{equation}
Note that this failure was not due to discretization, since, in the long wavelength limit, discrete models and  continuous partial differential equations produce the same result~\cite{Gomes19}. The result points to a possible change in dimension.

Reference \cite{Lima24} provides a geometric interpretation for this correction by arguing that, in the vicinity of the critical point, the effective activity of the system is confined to the fractal edge of spin clusters, which are characterized by a fractal dimension $d_R$. Within this hypothesis, the fractal dimension governing correlations is directly related to Fisher’s exponent, remaining distinct from the fractal dimension associated with the order parameter itself. Here, we rederive this result through a different route: relying on the randomness and macroscopic isotropy of the system, we assume that correlations on the percolation cluster admit a measure whose Fourier dimension equals its Hausdorff dimension, characteristic of Salem sets \cite{Salem1951,Fraser2020}. Furthermore, we employ a fractional version of the Gauss--Ostrogradsky theorem to directly relate Fisher’s exponent to the difference between the embedding spatial dimension and the fractal dimension of the correlations. However, as will be discussed in the following section, the use of the Riesz-Laplacian is not valid in the range $1\leq d \leq 3/2$, as it leads to a fractional derivative outside the bounds established in~\cite{Muslih10}. Therefore, these results are more general than those presented in \cite{Lima24}, as it relies neither on the Riesz--Laplacian Green's function construction nor on an analogy with electromagnetism. Consequently, this approach can be more naturally extended to other problems.


To validate and illustrate the proposal, we apply the framework discussed to the Ising model. The geometric formulation allows for a natural extension to continuous (non-integer) spatial dimensions $d$, remaining consistent below the upper critical dimension and providing a route to compute the Fisher exponent $\eta$ as a function of $d$. This geometric interpretation clarifies the role of $\eta$ and links fractal structure to universal scaling at criticality.

\section{ FRACTAL STRUCTURES AT CRITICALITY }

To understand the origin of the exponent $\eta$ in Eq. (\ref{G2}), we first note that the correlation functions represent a simple form of the Fluctuation--Dissipation Theorem (FDT) ( see reviews \cite{Nowak22,Oliveira19,GomesFilho25}), which is directly related to susceptibility \cite{Goldenfeld18}. This theorem is expected to hold in ergodic systems \cite{Costa03,Wen23}, and have a very subtle dimension dependence in growth systems~\cite{Kardar86,GomesFilho21,Wio10a,Anjos21,Rodriguez19,GomesFilho24}. Therefore, the fact that the simple solution of Eq. (\ref{G}) fails to reproduce the correct short-distance scaling, requiring the \textit{ad hoc} introduction of the exponent $\eta$, suggests that it might not remain valid at criticality. Since fractal structures emerge at $T = T_c$, the underlying geometry of the system becomes nontrivial, and this geometric transition could be responsible for the observed deviation.

The connection between the order parameter exponent $\beta$ and the fractal dimension of the ordered phase, $d_f$, was first proposed by Suzuki~\cite{Suzuki8} as
\begin{equation}
\label{dl}
d_f=d-\frac{\beta}{\nu}\;.
\end{equation}
At the critical point of the percolation model, this fractal structure corresponds to the infinite percolating cluster~\cite{Grimmett06,Cruz23}, while for general systems it is associated with the largest ordered cluster~\cite{Kroger00}. 

These scale-invariant fractal structures arise from fluctuations that span all length scales.~\cite{Kroger00,Grimmett06,Cruz23,Luis22}, this is why suppressing fluctuations in mean-field theories fails to reproduce the correct behavior. At the critical temperature $T_c$, the system exhibits a scale-invariant distribution of spin clusters, accompanied by the divergence of $\rho(T)$. The spatial variation of $G(r)$, and consequently of $\nabla^2 G$, becomes more relevant near the fractal boundaries of the clusters, precisely Eq. (\ref{G}) is nontrivial. Taking into account the long-range correlations and the limitations of Eq.~(\ref{G}), these observations suggest that the local operator $\nabla$ does not correctly represent how the correlations propagate through the complex geometries formed at criticality.

Since most spin flips are expected to occur along the fractal boundaries of clusters, and given the inherently nonlocal nature of fractional derivatives, we investigated whether reformulating Eq.~(\ref{G}) with a fractional Laplacian could yield the correct behavior. Consequently, we replace Eq.~(\ref{G}) by
\begin{equation}
	\label{G3}
	(-\nabla^2)^\zeta G(r)=\delta^{(d_R)}(r),
\end{equation}
where $d_R$ is a fractal dimension associated with the fractional derivative of order $\zeta$ applied to the correlation function, and the fractional $\delta$-function satisfies $\int \delta^{d_R}(\vec{r}-\vec{r'})f({ \vec{r'}})d^{d_R}\vec r'=f({\vec{r}})$, for any continuous function $f({\vec{r}})$. We use the $d_R$ to keep the notation consistent with our previous derivations, which relied on the definition of the Riesz-Laplacian, as in \cite{Lima25}. Equation~(\ref{G3}) can be viewed as a fractional generalization of the Poisson equation.

The mathematical framework of fractional calculus is not yet fully unified, with several distinct definitions of closely related operators having been proposed and explored in the literature. A review of its historical development and modern formulations can be found in \cite{Tarasov2021}. Since our goal here is to extract scaling behavior rather than to address formal mathematical subtleties, we adopt the simplest possible approach to analyze Eq.  (\ref{G3}). In doing so, we introduce an accessible notation that facilitates the discussion and provides useful intuition into the role of fractional calculus in the present context.

First, to integrate over the fractal support of the correlation function, we introduce an effective density function $g(\vec{r})$. The relation between the fractal measure denoted as $d^{d_R}r$ and the standard Euclidean (Lebesgue) measure $d^dr$ is established through this density, such that $d^{d_R}r = g(\vec{r})d^dr$. Therefore, writing the integral over the standard Euclidean measure $d^dr$ with this additional density serves as an effective representation of the integration over the fractal measure. Specifically, the relationship is defined such that:
\begin{equation}
\label{def_frac_int}
    \int^{L^d} g(\vec{r})d^dr=\int^{L^d}d^{d_R}r=L^{d_R}
\end{equation}
where $g(\vec{r})$ encodes the fractal support of the correlation function in our derivation, which can be interpreted as a kernel for the definition of the fractional integral \cite{Luchko2021}. Furthermore, a rescaling in the equation above shows how these measures are related under scale transformations. Thus, the density function must be homogeneous, satisfying:
\begin{equation}
	\label{homo}
	g(\lambda\vec{r}) = \lambda^{d_R-d}g(\vec{r}).
\end{equation}

The Fourier transform of a test function $\psi(\vec{x})$ on the fractal support $g(\vec{x})$ can be constructed as follows:
\begin{equation}
    \label{fract_furrier}
    \int \psi(\vec{r}) e^{i\vec{k}\cdot\vec{r}} d^{d_R}r=\int \psi(\vec{r}) e^{i\vec{k}\cdot\vec{r}} g(\vec{r})d^dr =\hat{\psi}(\vec{k}).
\end{equation}

At first glance, Eq. (\ref{fract_furrier}) appears to impose no explicit restriction on the wave vector $\vec{k}$. However, the dimension of the fractal support of the correlation function on the reciprocal space can never surpass the Hausdorff dimension of the original object. When these two dimensions coincide, the set is referred to as a Salem set \cite{Salem1951,Fraser2020}. Common situations in which these dimensions differ include deterministic fractals \cite{Fraser2023} and structures with strong anisotropy \cite{Orponen2013}, both of which may inhibit the decay of Fourier modes. In the present case, although cluster boundaries may exhibit multifractal behavior at local scales, the comparison with the Hausdorff dimension is governed by the asymptotic decay of the Fourier transform, which is a global property. Moreover, the underlying geometric object is a random fractal with isotropy at large scales in the critical regime. Therefore, according to \cite{Kahane1985}, it is reasonable to approximate it as exhibiting Salem-type behavior and to assume a Fourier-space density $\sigma(\vec{k})$ with the same homogeneity, $\sigma(\lambda \vec{k})=\lambda^{d_R-d}\sigma(\vec{k})$.



The inverse Fourier transform can be constructed following the same fractional prescription as in Eq. (\ref{fract_furrier})
\begin{equation} 
\label{fract_invfurrier}
	\begin{split}
    \frac{1}{(2\pi)^{d_R}}\int \hat{\psi}(\vec{k})e^{-i\vec{k}\cdot\vec{r}}d^{d_R}k=\frac{1}{(2\pi)^{d_R}}\int \hat{\psi}(\vec{k})e^{-i\vec{k}\cdot\vec{r}}\sigma(\vec{k}) d^{d}k=\psi(\vec{r}).
	\end{split}
\end{equation}

Considering Eq.~(\ref{G3}), a Fourier transform indicates that $\tilde{G}(k) \sim k^{-2\zeta}$. We can obtain the scaling of the correlation function in the real space by the inverse Fourrier transformation of $\tilde{G}(k)$. 
\begin{equation}
	\label{G4}
	G(r)\sim\int \frac{d^{d_R}k\exp(-i\vec{k}\cdot\vec{r})}{k^{2 \zeta}}.
\end{equation}

Using hyper spherical coordinates and substituting $y=kr$ together with the homogeneity of $\sigma$, we obtain
\begin{equation}
	G(r)\sim\int\frac{e^{-ikr\cos(\theta)}}{k^{2\zeta}}\sigma(\vec{k})k^{d-1}dk d\Omega,
\end{equation}
where $d\Omega$ is the solid angle.
By substituting $y = kr$ and using the homogeneity of $\sigma$, we find
\begin{equation}
\label{Gfrac}
	G(r)\sim r^{2\zeta-d_R}\left[\int y^{d-1-2\zeta} e^{-iy\cos(\theta)}\sigma(y)dyd\Omega \right],
\end{equation}
where $d\Omega$ is the solid angle. Since the integral is merely a constant, we obtain $G(r) \sim r^{2\zeta - d_R}$. Note that this scaling is consistent with Eq. (\ref{G3}), since $\nabla^{2\zeta}$ and $\delta^{(d_R)}(r)$ scale as $r^{-2\zeta}$ and $r^{-d_R}$, respectively. However, we do not explicitly rely on the Riesz--Laplacian here, as we aim to avoid the inconsistency mentioned before.

Requiring that the correlation function obtained in Eq.~(\ref{Gfrac}) reproduces the behavior of Eq.~(\ref{G2}) yields
\begin{equation}
	\label{dfs}
	[d-d_R]-2[1-\zeta]+\eta=0.
\end{equation}

Next, we consider the scaling of the fractional version of the Gauss--Ostrogradsky theorem
\begin{equation}
    \label{Gauss}
    \int_V \vec{\nabla}^\zeta\cdot(\vec{\nabla}^\zeta G(\vec{r}))d^dr\sim\oint_{\partial V}\vec{\nabla}^\zeta G(\vec{r})\cdot \hat{n} g(\vec{r})d^{d-1}r.
\end{equation}
A much more rigorous definition of the fractional Gauss--Ostrogradsky theorem can be found in \cite{Tarasov2021}. Rescaling $\vec{r} \to \lambda \vec{r}$ in Eq.~(\ref{Gauss}) leads to $\lambda^{d - d_R} = \lambda^{\zeta-1+2(d-d_R)}$, which yields
\begin{equation}
	\label{gauss4}
	d-d_R=1-\zeta.
\end{equation}
Combining Eqs.~(\ref{dfs}) and (\ref{gauss4}) gives
\begin{equation}
	\label{eta}
	\eta=d-d_R=1-\zeta.
\end{equation}

Therefore, reformulating Eq.~(\ref{G}) into its fractional form, Eq.~(\ref{G3}), leads to the result in Eq.~(\ref{eta}), which expresses the Fisher exponent as a function of the correlation fractal dimension. This interpretation suggests that the introduction of $\eta$ corresponds to a dimensional correction arising from the use of an integer (Euclidean) geometry in Eq.~(\ref{G}). Furthermore, although in the Riesz fractional Laplacian the parameters $d_R$ and $\zeta$ are constrained within the ranges $d - 1 \leq d_R \leq d$ and $ 1/2 \leq \zeta \leq 1$, their relation is not made explicit in ~\cite{Muslih10}. In contrast, Eq.~(\ref{eta}) provides a direct connection between these quantities, indicating that the deviation of the integer dimension $d$ to $d_R$ is directly related to the deviation of the derivative order $\zeta$ from unity.  To our knowledge, this important symmetry  has not been reported anywhere else. Equation~(\ref{eta}) supports the empirical observation that $\eta \ge 0$. However, note that Eq.~(\ref{eta}) implies a violation of $\zeta < 1/2 $ for $1\leq d \leq 3/2$,  as shown in Table~\ref{Table1}. By avoiding the explicit use of the Riesz Laplacian, the present formulation does not suffer from this inconsistency.
Finally, in possession of the Eq. (\ref{eta}), we just have use other very well known relations between the exponents (as presented in \cite{Lima24}) to relate $d_R$ to the fractal dimension of the ordered phase $d_f$~\cite{Suzuki8,Coniglio89} in Eq. (\ref{dl}), obtaining
\begin{equation}
\label{dl2}
d_R=2(d_f-1).
\end{equation}
 These relationships have been shown to be exact for known values of the Ising exponents in integer dimensions~\cite{Lima24}.

%


\section{Analytical approximations
and numerical results across continuous spatial dimensions}
\begin{table*}[t]
	
	\begin{tabular}{c|cccccc} \hline\hline
		$d$   & $\beta$       & $\nu$          & $\zeta$      & $d_f$         & $d_R$         & $\eta$ \\ \hline
		4     & 1/2           & 1/2            & 1            &  3            &  4            & 0\\
		3.75  & $0.458355(2)$ & $0.523405(15)$ & $0.99856(4)$ &  $2.87428(2)$ & $3.74856(4)$  & $0.001435(4)$\\
		3.5   & $0.41600(2)$  & $0.55215(2)$   & $0.99316(2)$ &  $2.74658(1)$ & $3.49316(2)$  & $0.00683(2)$ \\
		3.25  & $0.3723(1)$   & $0.5873(1)$    & $0.9821(1)$  &  $2.61608(6)$ & $3.2321(1)$   & $0.0178(1)$ \\
		3     & $0.3270(15)$  & $0.6310(15)$   & $0.963(3)$   &  $2.481(1)$   & $2.963(3)$    & $0.036(3)$ \\
		2.75  & $0.2800(30)$  & $0.686(3)$     & $0.933(5)$   &  $2.341(3)$   & $2.683(5)$    & $0.066(5)$ \\
		2.5   & $0.2305(3)$   & $0.758(5)$     & $0.891(3)$   &  $2.195(2)$   & $2.391(3)$    & $0.108(3)$ \\
		2.25  & $0.1790(25)$  & $0.857(6)$     & $0.832(4)$   &  $2.041(2)$   & $2.082(4)$    & $0.167(4)$ \\
		2     & 1/8           & 1              & 3/4          &  $15/8$       & $7/4$         &  1/4 \\
		1.875 & $0.097(3)$    & $1.10(1)$      & $0.698(4)$   &  $1.786(2)$   & $1.573(4)$    & $0.301(4)$  \\
		1.75  & $0.068(6)$    & $1.23(3)$      & $0.639(7)$   &  $1.695(3)$   & $1.389(7)$    & $0.361(7)$ \\
		1.65  & $0.045(10)$   & $1.37(7)$      & $0.58(1)$    &  $1.617(5)$   & $1.23(1)$     & $0.41(1)$ \\
		1.5   & $0.01(15)$    & $1.67(20)$     & $0.49(2)$    &  $1.494(8)$   & $0.99(2)$     & $0.51(2)$ \\
		1.375 & $-0.02(3)$    & $2.1(5)$       & $0.415(3)$   &  $1.38(1)$    & $0.79(3)$     & $0.61(3)$ \\
		1.25  & $-0.05(5)$    & $3.0(15)$      & $0.28(5)$    &  $1.26(2)$    & $0.53(5)$     & $0.71(5)$ \\
		1     & 0             & $ \infty $     & 0            &  1            & 0             & 1 \\ \hline\hline
	\end{tabular}
	\caption{Values of critical exponents for  the $d$ dimensional Ising model universality class. $\beta$, $\nu$ and $\eta^*$, from ref  \cite{Guillou87}, we use $\eta*$ to distinguish from $\eta$ obtained here from Eq.\ (\ref{eta}), $d_f$ from Eq. (\ref{dl}), $d_R$ from  Eq. (\ref{dl2}).}
	\label{Table1}
	
\end{table*}

For non-integer dimensions $d$, exact results are not available, and we therefore rely on numerical analysis to validate the above result. To obtain $d_R$, as in our previous calculations, one should not distinguish the percolating cluster from the others. Thus, one option is to apply the box-counting method to all spins aligned in the same direction as percolating cluster. In Ref.~\cite{Lima24}, we verified that this procedure indeed reproduces the expected value of $d_R$ in two dimensions. Furthermore, Eq. (\ref{dl2}) shows a profound relationship between our fractional approach and percolation theory.

Table~\ref{Table1} summarizes the values of $d_f$ and $d_R$ for the Ising model in different spatial dimensions, together with the corresponding critical exponents. The parameters $\beta$ and $\nu$ are taken from the literature, with  $\eta$ obtained through Eq.~(\ref{eta}). The exponents $d_f$ and $d_R$ are computed using Eqs.~(\ref{dl}) and (\ref{dl2}), respectively. For $d=4$, mean-field results are displayed; for $d=3$, the data are based on the estimates by Pelissetto and Vicari~\cite{Pelissetto02}; and for $d=2$, the exact exponents of the Ising model are used.

For non-integer dimensions ($1 < d < 4$), we employ the high-precision results obtained by Guillou and Justin~\cite{Guillou87} through resummation of the Wilson–Fisher $\epsilon$-expansion, using Borel transformation and conformal mapping techniques. These results reproduce the exact Ising exponents in $d=2$ and are consistent with those for $d=3$. In the range $1 < d < 2$, the less accurate estimates by Novotny~\cite{Novotny92} support these findings. For completeness, $d=1$ values are taken from the near-planar interface~\cite{Wallace79} and droplet~\cite{Bruce981,Bruce83} models.

The resulting $d_f$ values agree with the exact results by Coniglio~\cite{Coniglio89} for $d=2$ and $d=4$. As the dimensionality decreases, the $\eta$ values predicted by Eq.~(\ref{eta}) have a decreasing precision, as reported in Ref.~\cite{Guillou87} for low $d$. Nevertheless, the proposed approach reproduces the correct scaling even for non-integer dimensions, confirming its general applicability.  

\begin{figure}
	\centering
	\includegraphics[width=0.48\linewidth]{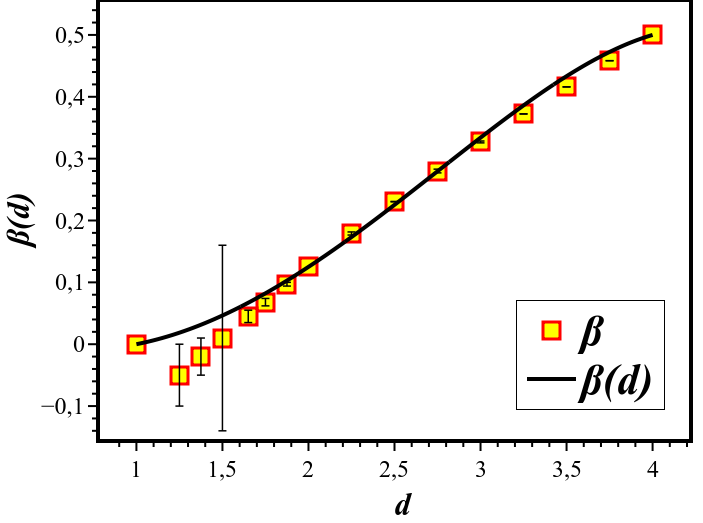}
    \includegraphics[width=0.48\linewidth]{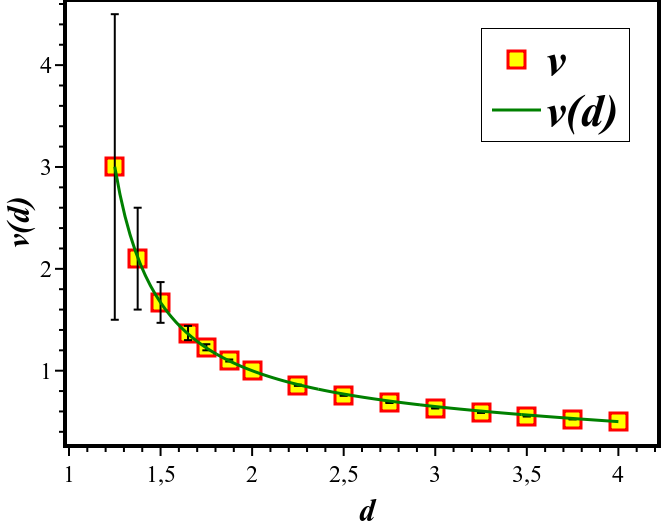}
    \includegraphics[width=0.48\linewidth]{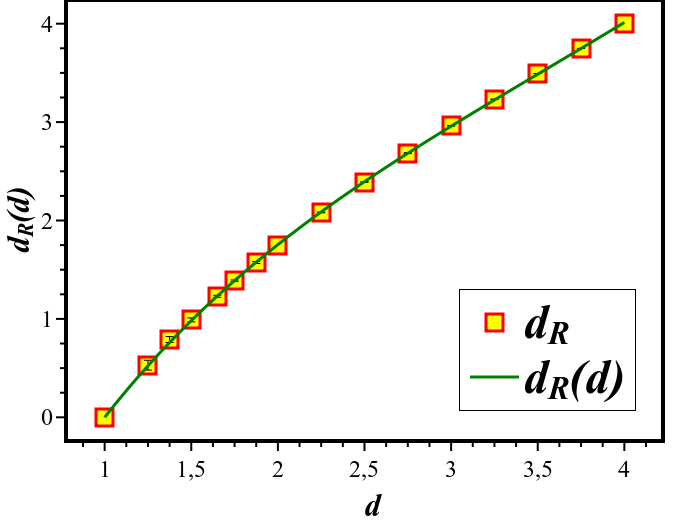}
    \includegraphics[width=0.48\linewidth]{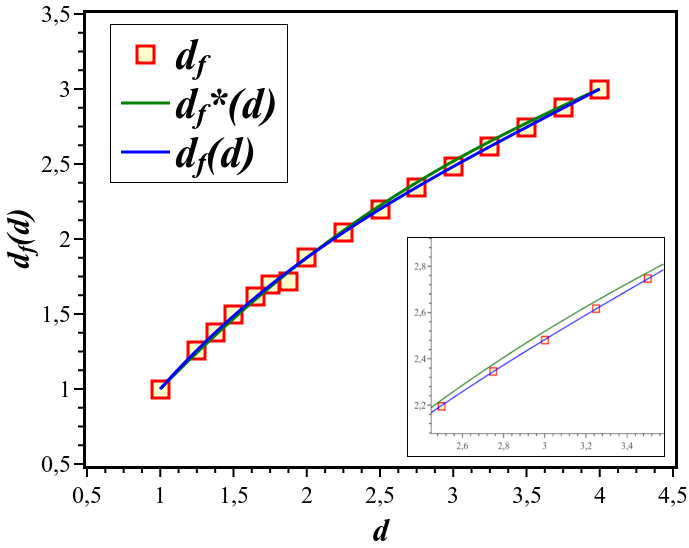}
	\caption{ Critical exponents as function of $d$ and their respective fitting curves. (a) exponent $\beta$ fitted by Eq. (\ref{curve_beta}), (b) $\nu$ fitted by Eq. (\ref{curve_beta}), (c) $d_R$ fitted by Eq. (\ref{curve_dr}) and (d) $d_f$ fitted by Eq.~(\ref{curve_df}).}
	\label{fig}
\end{figure}


We now propose expressions for the critical exponents as functions of the dimension, based on three guiding principles: first, they should be as simple as possible; second, they should reproduce the known exponents for $d=1,2$
$4$, which are captured by the term outside the brackets bellow; and finally, they should fit the data in Table~1, for which we introduce a small correction inside the brackets. Altogether
\begin{equation}
	\label{curve_beta}
	\beta(d) = \frac{d - 1}{10 - d}\left[1 - C_1 (4 - d)(2 - d)\right],
\end{equation}
and
\begin{equation}
	\label{curve_nu}
	\nu(d) = \frac{6 - d}{4(d - 1)}\left[1 - C_2 (4 - d)(2 - d)\right],
\end{equation}
where the constants $C_1 = 149/1000$ and $C_2 = 37/1000$ are determined by least-squares fitting within the range $1.5 \leq d \leq 4$. Here we ignore the region $d < 1.5$ given the much lower precision in the data, as mentioned before.

Likewise, the effective dimensions $d_R$ and $d_f$ can be expressed as
\begin{equation}
	\label{curve_dr}
	d_R(d) = \frac{56(d - 1)}{5d + 22}\left[1 - \frac{2152(4 - d)(d - 2)}{100000}\right],
\end{equation}
and, from Eq.~(\ref{dl2}),
\begin{equation}
	\label{curve_df}
	d_f(d) =1+\frac{d_R}{2}= 1 + \frac{28(d - 1)}{5d + 22}\left[1 - \frac{2152(4 - d)(d - 2)}{100000}\right]
\end{equation}
 which, starting from Eq. (\ref{dl}), should be compared with the expected relationship
\begin{equation}
\label{dfestrela}
d_{f}^*(d)=d-\frac{\beta(d)}{\nu(d)},    
\end{equation}
where $\beta(d)$ is from Eq. (\ref{curve_beta}) and $\nu(d)$ is from (\ref{curve_nu}). 

Figure~\ref{fig} presents the critical exponents $\beta$ and $\nu$, as well as the fractal dimensions $d_R$ and $d_f$. In each panel, the data points from Table~\ref{Table1} are shown together with the corresponding fits obtained from Eqs.~(\ref{curve_beta})--(\ref{curve_df}). Overall, the fitted curves reproduce the data remarkably well, including in the region $1 < d < 1.5$, which was not included in the fitting procedure. The most noticeable deviation occurs for the exponent $\beta$, where the nonphysical negative value reported in the table is not captured by the fit. The fitted curve is a monotonically increasing function starting from zero, which appears to be physically more reasonable. 

The overall consistency between the fitted expressions and the tabulated values supports the view that Eqs.~(\ref{curve_beta}), (\ref{curve_nu}), (\ref{curve_dr}), and (\ref{curve_df}) provide reliable empirical approximations for the critical exponents $\beta$ and $\nu$, as well as for the fractal dimensions $d_R$ and $d_f$, across the continuous range $1 \leq d \leq 4$. Finally, we compare these results with the expected scaling relation. In the inset of Fig.~\ref{fig}d, we observe very good agreement between the curves, further supporting the internal consistency of Eqs.~(\ref{curve_beta})--(\ref{curve_df}).   A repository for these dates are in \cite{Carrasco260r}.


\section{Concluding Remarks}

In this work, we have revisited the relationship between the correlation function and the underlying fractal structure of spin clusters in the Ising model at criticality. Our earlier formulations relied directly on the use of the Riesz--Laplacian due to its well-known Green's function and an analogy with electromagnetism. However, the Riesz--Laplacian approach was not valid in the problematic range $1 \leq d \leq 3/2$, as it required $\zeta \leq 1/2$, falling outside the bounds established in~\cite{Muslih10}. To overcome this, we have reexamined the problem by introducing integrations over a fractal support and utilizing the fractional Fourier transforms. Assuming that the critical fractal object maintains the same effective dimensionality in reciprocal space, a property of Salem sets, we recovered the scaling of the correlation function via Fourier analysis. Furthermore, by employing a fractional version of the Gauss--Ostrogradsky theorem, we derived Fisher’s exponent as the difference between the embedding spatial dimension and the correlation fractal dimension. Ultimately, this generalized formulation not only validates and reinforces previous results by circumventing the $\zeta \leq 1/2$ inconsistency, but it also offers a more systematic and conceptually transparent approach. This highlights the utility of fractional calculus in analyzing systems with long-range correlations and related critical phenomena.

The central outcome of this approach is the exact relation given by Eq.~(\ref{eta}), which identifies the Fisher exponent $\eta$ as the difference between the Euclidean dimension $d$ and the correlation fractal dimension $d_R$. This result reveals that $\eta$ directly quantifies the deviation of the geometry of correlations from that of the embedding Euclidean space. In parallel, the relation between $d_R$ and the fractal dimension of the ordered phase $d_f$, given by Eq.~(\ref{dl2}), establishes a direct link between geometric and thermodynamic descriptions at criticality. Moreover, the good agreement between the analytical predictions and tabulated critical exponents across different spatial dimensions provides strong evidence that Eqs.~(\ref{eta}) and~(\ref{dl2}) hold universally.

Finally, exploiting the fact that our formulation naturally extends to continuously varying spatial dimensions, we obtain explicit expressions for the critical exponents $\beta$ and $\nu$, as well as for the fractal dimensions $d_f$ and $d_R$ as a function of $d$, given in Eqs.~(\ref{curve_beta})--(\ref{curve_df}) and illustrated in Figure ~\ref{fig}. These expressions yield accurate numerical approximations for all dimensions in the range $1 \leq d \leq 4$.

This geometric interpretation offers an alternative perspective to the traditional notion of a critical upper dimension. As $d$ increases, the difference $d - d_R$ gradually disappears, and the developed fractal mean-field approach smoothly converges to traditional Euclidean mean-field behavior. Therefore, the usual mean-field regime emerges naturally as the limit in which the correlation geometry becomes purely Euclidean, i.e., $d = d_c = d_R$.

Finally, the formulation presented here is not restricted to the Ising model or to equilibrium transitions. Since Eq.~(\ref{eta}) was derived for a general continuous order parameter $\psi(\vec{r})$, it can be extended to systems with quenched disorder or non-equilibrium dynamics. Exploring these directions, particularly in the context of disordered systems~\cite{Vainstein05} and  dynamic phase transitions~\cite{Pinto16,Pinto17,Santos24}, where the autocorrelation function (\ref{G0}) will need to include the temporal dependence~\cite{Henkel26}, may open new avenues for understanding the interplay between fractal geometry, universality, and critical phenomena. Recent research directly connected with this work such as: Schrödinger invariance in the voter model~\cite{Henkel26}, on conformal scaling and critical phenomena~\cite{Weberszpil26} and Fisher curvature scaling at critical points~\cite{Zhuravlev26} shows that this is a promising field with many questions to be answered. 

All data used in this work are publicly available and can be accessed at  \cite{Carrasco260r}.

{\bf Acknowledgments -} This work was supported by 
the Conselho Nacional de Desenvolvimento Cient\'{i}fico e Tecnol\'{o}gico (CNPq), Grant No.  CNPq-01300.008811/2022-51, 303119/2022-5  and Funda\c{c}\~ao de Apoio a Pesquisa do Distrito Federal (FAPDF), Grant No.\ 00193-00001817/2023-43.

\bibliography{sn-bibliography1}

\end{document}